\documentclass[english,preprint,nofootinbib]{revtex4-2}
\usepackage[utf8]{inputenc}
\usepackage{xcolor}
\usepackage{pdfcolmk}
\usepackage{float}
\usepackage{amsmath}
\usepackage{graphicx}
\PassOptionsToPackage{normalem}{ulem}
\usepackage{ulem}

\makeatletter

\DeclareTextSymbolDefault{\textquotedbl}{T1}
\providecolor{lyxadded}{rgb}{0,0,1}
\providecolor{lyxdeleted}{rgb}{1,0,0}

\DeclareRobustCommand{\lyxsout}[1]{\ifx\\#1\else\sout{#1}\fi}

\usepackage{babel}
\newcommand{\degree}{ ^{\circ}}
\makeatother

\begin{document}
\title{  Irreversibility of heat transport and optical non-coherence: an experimental demonstration of the analogy }
\author{Aleksandr Meilakhs}
\email[A.\,P. Meilakhs: ]{iestfi@gmail.com}
\affiliation{Departamento de Física de Materia Condensada, GIYA, CAC-CNEA, Av. Gral. Paz 1499, San Martín, Pcia. Buenos Aires, Argentina}
\affiliation{Instituto de Nanociencia y Nanotecnología, INN-CONICET-CNEA}
\author{Claudio Pastorino}
\affiliation{Departamento de Física de Materia Condensada, GIYA, CAC-CNEA, Av. Gral. Paz 1499, San Martín, Pcia. Buenos Aires, Argentina}
\affiliation{Instituto de Nanociencia y Nanotecnología, INN-CONICET-CNEA}
\author{Miguel Larotonda}
\affiliation{Laboratorio de Optica Cuántica, DEILAP, UNIDEF (CITEDEF-CONICET), Buenos Aires, Argentina}
\affiliation{Departamento de Física, Facultad de Ciencias Exactas y Naturales, UBA, Ciudad de Buenos Aires, Argentina}

\date{\today}


\begin{abstract}

The arrow of time problem remains one of the most intriguing questions of
modern physics. We investigate one particular example of this problem: the
irreversibility of heat transfer through an interface between two
materials. This special case turned out to be much simpler than that of
transport phenomena in bulk materials. We start by exploring a mechanism that turns equations of reversible,
unitary transmission of wave-particles through the interface into
irreversible equations, describing the distribution functions at the
interface. This mechanism, non-coherence, is treated in an analogy
with wave optics: no real waves are absolutely monochromatic and thus have finite temporal coherence. We find an experimental setup 
to prove the suggested mechanism and calculate the optimal
parameters to observe the phenomenon.  We conduct an experiment whose results fully agree with the
predictions of our theory. In the final discussion, we revisit the
connection with heat transport through a material interface and show that only the
suggested mechanism can explain the irreversibility of interfacial transport.

\end{abstract}

\maketitle
\section{Introduction}

One of the most intriguing problems in modern physics is the irreversibility
of macroscopic processes. As it was first noticed in the context of
thermodynamics, the transport of heat from a hot body to a cold one
is gives rise to entropy production. Since entropy can only
grow and never decrease such processes are irreversible \cite{Landau1994,Kondepudi2015}.
Later, it was found that the Boltzmann equation, along with other
kinetic equations governing the behavior of large ensembles of particles
is irreversible \cite{Landau2013}. This highlights a drastic contrast
with the equations describing individual particles, whether in a
classical or a quantum framework.

The most widespread approach for the derivation of the Boltzmann equation
from classical mechanics is the so-called BBGKY hierarchy \cite{Harris2004}.
A serious drawback of this method is the postulate of molecular chaos,
as a model for the two-particle distribution function, whose fundamentals
are not properly explained. Another approach to explain entropy growth
postulates an initial low entropy state for every process. Equivalently
it postulates low-entropy initial states of the universe \cite{CARROLL2005,Wald2006},
which is an \emph{ad hoc} hypothesis \cite{Cohen1960} and hence,
not completely satisfactory. Some authors \cite{Maccone2009} propose
modifications of the physical laws, to explicitly incorporate irreversibility.

Most modern approaches propose the emergence of irreversibility from
a time-asymmetric process, which is an inherent feature of quantum
physics \cite{Ghirardi1986}. The most studied of such mechanisms
is decoherence \cite{Schulman1997,Joos2003}. Decoherence theory divides changes occurring in a quantum system into two categories: normal evolution, described by the Schrödinger equation, and measurement. The latter is then associated with the process of decoherence. Of these two parts, it is decoherence that is considered irreversible \cite{Schulman1997}. According to this theory, decoherence arises due to the interaction of the quantum system with its environment.

Quantum thermodynamics theory also attributes irreversibility to phase
fluctuations (non-coherence) that result from an interaction of the system with the environment
\cite{ breuer2002openquant}. This approach originated with the introduction of
the Lindblad master equation \cite{lindblad1976generators}, which describes the
dynamics of a subsystem in contact with a heat bath. This theory has been
applied to various models in quantum thermodynamics
\cite{deutsch2010thermodynamic, joshi2022probing}. For example, it has been shown that
systems in contact with a heat bath show a behaviour similar to models that are
used in classical statistical mechanics, such as the microcanonical ensemble
\cite{dymarsky2018subsystem} and the Boltzmann equation
\cite{rigol2008thermalization}.

In one of our previous works \cite{meilakhs2024transmission}, the
problem of irreversibility of heat transport through an interface
between two media was addressed. It is known that a temperature jump
occurs at the interface when heat flows through two different media.
The proportionality coefficient between the heat flux and the temperature
jump is usually called the Kapitza resistance.

Currently, the Kapitza resistance has become a wide area of research,
not only for its theoretical interest, but also for its technological
relevance in a wide variety of thermal management applications, see Ref. \cite{nandihalli2018effect}. Some
works investigate the dynamics of the crystal lattice at the interface
through computer simulations \cite{Saeaeskilahti2014,Yang2015,Bi2016,Alkurdi2017,Kakodkar2017,Huang2018},
while there is also a great number of works with analytical approaches
\cite{Young1989,Zhang2011,Meilakhs2016}. Other studies deal with
phonon kinetics at the interface with Boltzmann theory \cite{Majumdar2004,Merabia2012,Alaili2019,Varnavides2019}
or the nonequilibrium Green's function method \cite{Tian2012,Wang2007,Wang2008}.
Measurements of values of Kapitza resistance for different pairs of
materials were also performed \cite{Costescu2003,Ye2017, nandihalli2018effect}. Additionally,
the problem of Kapitza resistance was studied
in a broader context than phonon transport, such as electron transport
between semiconductors \cite{Meilakhs2023} and molecular transport
at a liquid-vapor interface \cite{Pastorino2022}.

After the discovery of the Kapitza resistance, it was very soon realized
that the temperature jump is due to the reflection of phonons at the
interface \cite{Khal}. Since phonon reflection at a solid-solid interface
is a linear process, the Kapitza resistance effect can be used as
a testbed to shed light on the mechanism that leads to irreversibility.
In contrast with the time-dependent and highly non-linear Boltzmann
equation, its analog at the interface between two materials, the matching
equation for the distribution functions, \cite{Meilakhs2015,Meilakhs2021}
is linear and static.

In Ref. \cite{meilakhs2024transmission} the interfacial phenomenom of Kapitza
resistance was related to the concept of non-coherence.  It was suggested that
if particles incident on the interface from different sides are
phase-correlated (coherent regime) then their transmission through the
interface occurs reversibly. On the contrary, if there is no phase correlation
between particles at both sides of the interface (non-coherent regime), which
is the naturally occurring situation, the transmission becomes irreversible. In
that work, a thought experiment was proposed to establish this
connection between non-coherence and ireversibility. In the present manuscript, we establish all the important aspects of an actual experiment. We use an optic setup, where the interaction of light with a dielectric interface simulates both a reversible and a non-reversible transport regime. For this optical simulator, we employ two light sources to produce either a coherent or a non-coherent beam, at both sides of the interface. In order to obtain an interaction region at the surface we chose the angle of incidence that matches the propagation distance of the light at both sides of the interface in the longitudinal projection (i.e. the direction of the interface). Taking into account losses in the reflecting surfaces, we obtain results that show the validity of the proposed connection between coherence and reversibility.

We point out that theories that suggest interaction with the environment as
a source of phase fluctuations and hence irreversibility, such as decoherence
theory or quantum thermodynamics, can not explain the irreversibility of the
interfacial transport. Transmission or reflection at the interface happens
instantaneously (with great precision), which prevents the interaction with the
heat bath. To achieve irreversible behavior the photons that are incident at the interface from different
sides should already be phase-uncorrelated at the moment of incidence. We will
discuss this matter in more detail in the text of the manuscript.

 The time irreversibility 
shown by non-coherent transmission through an interface is put to test with 
photons incident on an interface between two media and a two-mirror 
arrangement, in a conceptually simple optical experiment that implements 
a Mach-Zehnder interferometer \cite{hariharan2010basics,soler1997multiple,prikhodko2024effects}: 
an input light beam is split into two according to the Fresnel reflection and 
refraction coefficients of an air-glass interface. Different outcomes arise 
from the interaction of these two light beams at the same interface, determined 
by the coherence properties of the initial beam. With a highly coherent input 
beam, complete recombination into a single output beam can be achieved. In 
contrast, if a non-coherent source is used, a constant fraction of the light 
is transmitted and reflected at the interface, respectively, regardless of the 
prior evolution of the interacting beams\footnote{We refer to light sources as 
highly coherent and non-coherent when their coherence lengths are much longer and 
much shorter, respectively, than the optical path difference between the two 
interacting beams \cite{born2013principles}.}. 

In section \ref{sec:Theory} we present the theoretical framework within
the context of light transport, an ideal experiment and a feasible
experimental realization. In section \ref{sec:The-experiment}, the
 experimental setup is described in detail. We devote
section \ref{sec:Results} to present the results and a final discussion
is provided in section \ref{sec:Discussion}.

\section{Theory\label{sec:Theory}}

\subsection{Preambula}

\begin{figure}[H]
\centering \includegraphics[width=0.45\textwidth]{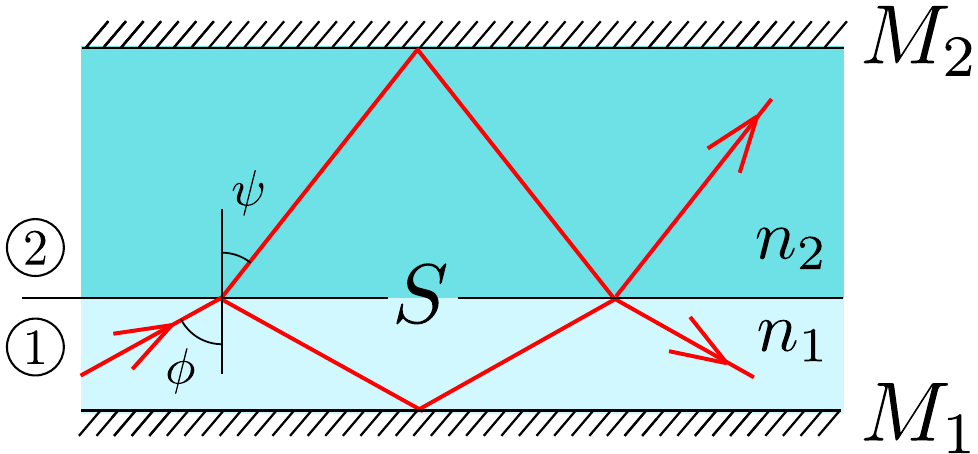}
\caption{General scheme for the study of coherence conditions that lead to
reversible and irreversible processes of photons at an
interfacial boundary. Light incident on an interface $S$ between two optical
media of different refraction indices $n_{1}$ and $n_{2}$ gets partially
reflected. The refracted and reflected beams are re-routed back to
the interface by means of the mirrors $M_{1}$ and $M_{2}$, where they interact in different
    ways depending on the coherence condition of the light.}
\label{fig:genscheme} 
\end{figure}

We studied the light transport through an interface of two materials
under two different coherence conditions of the input field, as sketched
in Fig. \ref{fig:genscheme}. A light beam is split at a partially
reflecting interface $S$. The reflected part of the light travels
through one of the optical media and bounces at a high reflectivity
surface $M_{1}$, which is parallel to the interface. Concurrently,
the refracted (transmitted) part of the light travels through a different
medium and hits a second highly reflective surface $M_{2}$. After
their respective reflections, both beams are recombined at the interface.
At this point, the coherence condition of the light imposes a limit
on the reversibility of the process, \emph{i.e.}, whether the light
can get totally or partially recombined into a single output beam.
Since the materials have different refractive indices, their relative
thicknesses must be adjusted to compensate for the differing propagation
lengths caused by the refraction and reflection angles. Under such condition, 
the second point of interaction is shared by both beams at the interface.

We consider linearly polarized waves, with a polarization perpendicular
to the plane of incidence. If the light is incident on the interface
from one side, the amplitudes of transmitted (refracted) and reflected
light,  with polarization perpendicular to the plane of incidence (s-polarization), are given by the Fresnel formula \cite{Hecht}: 
\begin{align}
r_{S}=\frac{\sin(\psi-\phi)}{\sin(\phi+\psi)},\ \ \ t_{S}=\frac{2\sin\psi\cos\phi}{\sin(\phi+\psi)}\label{Fresnel}
\end{align}
Here, $\phi$ and $\psi$ are  the angles between the incident and transmitted propagation directions and the axis perpendicular
to the interface, respectively (Fig. \ref{fig:genscheme}). The interface is a boundary
between two optical materials of refractive indices $n_{1}$ and $n_{2}$.
The Snell's law of refraction imposes the condition for the incident
and refracted angles $\phi$ and $\psi$: $n_{1}\sin\phi=n_{2}\sin\psi$.
For the particular condition of an interface between air and optical
glass, $n_{1}\approx1$ and $n_{2}=n$, hence $\sin\phi=n\sin\psi$.

If light is incident at the interface from both sides with amplitudes
$A_{I1}$ and $A_{I2}$, and the angles of incidence are matched by
Snell's law, the amplitudes of departing waves $A_{D1}$, $A_{D2}$,
can be found by 
\begin{align}
\begin{pmatrix}A_{D1}\\
A_{D2}
\end{pmatrix}=\mathcal{A}\begin{pmatrix}A_{I1}\\
A_{I2}
\end{pmatrix}.\label{Transformation}
\end{align}
where $\mathcal{A}$ is the matrix of transfomation of amplitudes
defined by 
\begin{align}
\mathcal{A}=\frac{1}{\sin(\phi+\psi)}\begin{pmatrix}\sin(\psi-\phi) & 2\sin\phi\cos\psi\\
2\sin\psi\cos\phi & \sin(\phi-\psi)
\end{pmatrix}.\label{Transformatrix}
\end{align}
The values of the columns in matrix $\mathcal{A}$ are both obtained
from equation \eqref{Fresnel}, with the second column assuming that
the light is incident from the right direction.

In \cite{meilakhs2024transmission} it is shown that, if we normalize
the amplitudes such that its squares at a given side and direction
are equal to the intensity $I$  at a given side and direction, i.e. $|A|^{2}=I$,
the transformation matrix becomes unitary and it is given by the following
expression
\begin{align}
\mathcal{U}=\frac{1}{\sin(\phi+\psi)}\begin{pmatrix}\sin(\psi-\phi) & 2\sqrt{\sin\psi\cos\psi\sin\phi\cos\phi}\\
2\sqrt{\sin\psi\cos\psi\sin\phi\cos\phi} & \sin(\phi-\psi)
\end{pmatrix}.
\label{Unitary}
\end{align}
 Throughout the remainder of the text, we will use
a system of units in which squares of amplitudes represent intensities,
and the transformations of amplitudes are given by unitary matrices.

It is well known that unitary matrices form a group, and for any unitary
matrix, there exists an inverse matrix, which is also unitary. That is the
mathematical essence of what we mean by saying transformations are reversible.

In the presented notation, the problem can be written as follows.
We initially have a light beam at only one side of the interface.
Let us define its amplitude as one: 
\begin{align}
\begin{pmatrix}A_{I1}\\
A_{I2}
\end{pmatrix}=\begin{pmatrix}1\\
0
\end{pmatrix}.
\end{align}

Upon incidence at the interface, it undergoes a transformation given
by matrix $\mathcal{U}$ (Eq. \ref{Unitary}). Then the light is split
into two parts, each of which evolves freely so that, before the next
incidence on the interface, their phases will undergo the transformation
$\exp(ikl)$. Here $k$ is the corresponding wave vector and $l$ is
the distance that light travels between two instances of incidence
at the interface. When the beams meet at the interface for the second
time, their amplitudes are again transformed by $\mathcal{U}$ (see
Fig. \ref{fig:genscheme}). So we have the following final amplitudes:
\begin{align}
\begin{pmatrix}A_{F1}\\
A_{F2}
\end{pmatrix}=\mathcal{U}\begin{pmatrix}e^{ik_{1}l_{1}} & 0\\
0 & e^{ik_{2}l_{2}}
\end{pmatrix}\mathcal{U}\begin{pmatrix}1\\
0
\end{pmatrix}.\label{Coherent1}
\end{align}
The intensities are given by the squares of the amplitudes $|A_{F1}|^{2},|A_{F2}|^{2}$.
These intensities are ultimately measured in our experiment.

If we multiply both amplitudes on the same quantity with module 1,
the resulting intensities would not change. Therefore, the important
quantity that characterizes the matrix of free propagation is $\exp i(k_{1}l_{1}-k_{2}l_{2})$,
\emph{i.e,} the phase difference between the two paths of propagation.

Since $\mathcal{U},$ given by Eq. \eqref{Unitary}, is symmetric
then $\mathcal{U}^{\dag}=\mathcal{U}$. Here the dagger symbol represents
the Hermitian conjugation. We recall that the defining property of
unitary matrices is $\mathcal{U}^{\dag}\mathcal{U}=\mathcal{E}$,
where $\mathcal{E}$ is the identity matrix. Combining these two properties
we find that $\mathcal{U}^{2}=\mathcal{E}$.

If $k_{1}l_{1}-k_{2}l_{2}=2\pi N$, for any natural value of $N$
($N\in\mathbf{N}$), the matrix of free propagation is equivalent
to the identity matrix. This means that the product of the three matrices
in the expression \eqref{Coherent1} is equal to the identity matrix:
$\mathcal{U}\mathcal{E}\mathcal{U}=\mathcal{E}$. Therefore,the
system that went through these series of transformations will come
back to its initial state. More concretely, the two beams initially
split at the interface, would merge back into a single one. That is
the physical essence of the transformation being reversible.

This reasoning is valid when the light is completely coherent, or in other words, 
completely monochromatic. However, in reality, every light source has a finite bandwidth 
$\Delta \omega$ and thus a finite spectrum of wave vectors $\Delta k =  \Delta \omega/c$. 
If  $\Delta k(l_1 - l_2) \ll 1$, we can neglect the finite bandwidth and treat the 
light as completely coherent. 

Consider the opposite, non-coherent case $\Delta k(l_1 - l_2) \gg 1$. We denote the phase 
difference $\Delta\phi = \Delta k(l_1 - l_2)$ and to obtain the total intensity we average 
over all phase differences
\begin{align}
I =\frac{1}{\Delta\phi} \int \limits_0^{\Delta\phi} d\phi |A(\phi )|^2.
\label{fullint}
\end{align}
By replacing the values of amplitudes from equation \eqref{Coherent1} into this expression, 
we obtain for the intensity on the first side of the interface:
\begin{align}
I_{F1} = \frac{1}{\Delta\phi} \int \limits_0^{\Delta\phi} d\phi \left( |U_{11}|^2 |U_{11}|^2 + |U_{12}|^2 |U_{21}|^2 + \overline U_{11} U_{12} \overline U_{12} U_{21} e^{i\phi} + 
U_{11} \overline U_{12} U_{12} \overline U_{21} e^{- i\phi} \right). 
\label{Intencity}
\end{align}
We can write an analogous equation for the second side. If $\Delta\phi$ is large, integration 
of $\exp(i\phi)$ over $\Delta\phi$ yields zero. Considering this, we can write
\begin{align}
I_{F1} = |U_{11}|^2 |U_{11}|^2 + |U_{12}|^2 |U_{21}|^2 \nonumber \\
 I_{F2} = |U_{21}|^2 |U_{11}|^2 + |U_{22}|^2 |U_{21}|^2.
\label{IntFin}
\end{align}
We introduce the matrix
\begin{align}
\mathcal{T}= \begin{pmatrix}|U_{11}|^2 &  |U_{12}|^2 \\
|U_{21}|^2 &  |U_{22}|^2
\end{pmatrix}.\label{Bistoch1}
\end{align}
which, for $\mathcal{U}$ given by expression \eqref{Unitary}, it can be written as
\begin{align}
\mathcal{T}=\frac{1}{\sin^{2}(\phi+\psi)}\begin{pmatrix}\sin^{2}(\psi-\phi) & 4\sin\psi\cos\phi\sin\phi\cos\psi\\
4\sin\psi\cos\phi\sin\phi\cos\psi & \sin^{2}(\phi-\psi)
\end{pmatrix}.
\label{Bistoch}
\end{align}
Therefore, using this notation, equation \eqref{IntFin} can be compactly written as
\begin{align}
\begin{pmatrix}I_{F1}\\
I_{F2}
\end{pmatrix}=\mathcal{T}^{2}\begin{pmatrix}1\\
0
\end{pmatrix}.
\label{Non-coherent1}
\end{align}
We conclude that the non-coherent case may be described in terms of intensities and their 
transformations by $\mathcal{T}$-matrices, rather than with amplitudes and their 
transformations by unitary matrices $\mathcal{U}$.

In unitary matrices, the sum of the squares of the absolute values of each element of a 
given row (or column) is equal to one. Since $\mathcal{T}$-matrices are obtained from unitary 
matrices by taking squares of absolute values of each element, the sum of all elements of each 
row (and column) is equal to one. Matrices of this type are called bistochastic. Such
matrices lack inverses within this class of matrices. That is the
mathematical essence of the phenomenon that is to be demonstrated:
the correspondence of loss of reversibility with the loss of coherence.

From expression \eqref{Non-coherent1} we can see that the intensities are not dependent on 
the optical paths, they are strictly greater than zero, and lower than one. It means that non-coherent
beams cannot be combined back into a single beam, which is the physical
manifestation of the irreversibility of non-coherent transformations.

\subsection{Scheme of the ideal experiment}

Our goal is to distinguish the coherent regime from the non-coherent
one. The most convenient way of observing the coherence is by measuring
the visibility of the optical fringes, defined by the expression 
\begin{equation}
\nu=\frac{I_{{\rm {max}}}-I_{{\rm {min}}}}{I_{{\rm {max}}}+I_{{\rm {min}}}}.\label{Visibility}
\end{equation}

Experimentally, these maximum and minimum intensity values can be
obtained by slightly varying the size of the air gap.

We will show experimentally that the visibility $\nu$ is zero in
a non-coherent, irreversible regime, whereas it can reach values close
to unity for a reversible, coherent case. In real conditions the visibility
of a coherent light beam may be ultimately limited by experimental
factors such as a distorted spatial mode, or scattering in the
reflecting surfaces. Before reaching this condition, several aspects
of the particular setup must be considered. In what follows, we describe
the experimental conditions needed to obtain the highest possible
degree of reversibility (maximum visibility) with the available light
source.

In an ``ideal'' experiment, a light beam incident at a boundary
between two isotropic media is split into a
reflected
beam and a transmitted one. These two beams are totally reflected
on parallel surfaces and re-routed to a common point at the boundary
surface. In order to maximize visibility, the amplitudes of both beams
must be equal. These amplitudes are given by the reflection and transmission
coefficients that can be obtained from Eq. \eqref{Fresnel}

Let us denote the elements of the unitary transformation matrix $\mathcal{U}$
given by the expression in Eq. \eqref{Unitary} as 
\begin{align}
\mathcal{U}=\begin{pmatrix}a & b\\
b & -a
\end{pmatrix}.\label{Uny1}
\end{align}
The minimum intensity is obtained when the waves are incident on the
interface exactly in antiphase. Therefore, the full transformation
matrix
\eqref{Coherent1} reads: 
\begin{align}
\begin{pmatrix}a & b\\
b & -a
\end{pmatrix}\begin{pmatrix}1 & 0\\
0 & -1
\end{pmatrix}\begin{pmatrix}a & b\\
b & -a
\end{pmatrix}=\begin{pmatrix}a^{2}-b^{2} & 2ab\\
2ab & -a^{2}+b^{2}
\end{pmatrix}.\label{Antyphase}
\end{align}
The minimum is then obtained for $a^{2}=b^{2}$, which means that
the intensities are split equally in both directions.

This results in the expression $\sin^{2}(\psi-\phi)=4\sin\psi\cos\psi\sin\phi\cos\phi$.
We use the substitution $\sin\psi=n_{1}\sin\phi/n_{2}$ and, after
some manipulations, we find 
\begin{align}
\sin\phi=\sqrt{\frac{n_{2}^{2}+n_{1}^{2}-\sqrt{9/8}(n_{2}^{2}-n_{1}^{2})}{2}}.\label{Optimal}
\end{align}
This gives the incidence angle for a balanced reflection and refraction
amplitudes, as a function of the refractive indices on both sides
of the interface.

\subsection{Scheme of the real experiment \label{realexp}}

However, since mirrors do not have perfect reflectivities, part of
the light is absorbed by them. In order to include these losses, a
more realistic transformation of the amplitudes between incident beams
on the interface must include the multipliers that affect the absolute
values of the amplitudes: 
\begin{align}
\begin{pmatrix}A_{F1}\\
A_{F2}
\end{pmatrix}=\mathcal{U}\begin{pmatrix}r_{1}e^{ik_{1}l_{1}} & 0\\
0 & r_{2}e^{ik_{2}l_{2}}
\end{pmatrix}\mathcal{U}\begin{pmatrix}1\\
0
\end{pmatrix}.\label{Coherent2}
\end{align}
and for a non-coherent transformation, we obtain 
\begin{align}
\begin{pmatrix}I_{F1}\\
I_{F2}
\end{pmatrix}=\mathcal{T}\begin{pmatrix}R_{1} & 0\\
0 & R_{2}
\end{pmatrix}\mathcal{T}\begin{pmatrix}1\\
0
\end{pmatrix}.\label{Non-coherent2}
\end{align}
Here,$r_{1}=\sqrt{R_{1}}$, $r_{2}=\sqrt{R_{2}}$, and the parameters
$R_{1}$ and $R_{2}$ must be determined experimentally.

The conditions for maximal visibility is analogous to that of Eq.
\eqref{Antyphase}: 
\begin{align}
\begin{pmatrix}a & b\\
b & -a
\end{pmatrix}\begin{pmatrix}r_{1} & 0\\
0 & -r_{2}
\end{pmatrix}\begin{pmatrix}a & b\\
b & -a
\end{pmatrix}=\begin{pmatrix}a^{2}r_{1}-b^{2}r_{2} & ab(r_{1}+r_{2})\\
2ab(r_{1}+r_{2}) & -a^{2}r_{1}+b^{2}r_{2}
\end{pmatrix}.\label{Antiphase2}
\end{align}

We find that the condition of maximum visibility , which is $a^{2}r_{1}-b^{2}r_{2}$ in
this case, has the same physical meaning as that of the ideal case:
at the recombination of beams, intensities on both sides should be
equal.

Taking into account the losses due to imperfect reflection at the mirrors, it must 
be noticed that we will not obtain the same intensity at the output and the input. Therefore,
we do not aim for the complete reversibility of the whole experiment. What we
want to demonstrate is the reversibility of the process of transmission/reflection
of light at the interface between the air and the glass. We achieve
this by showing that all the light that is left in the system after
the reflections at the mirrors, is merged into a single beam.

\section{The experiment\label{sec:The-experiment}}

\subsection{Experimental Setup}

The light source that we define as a ``coherent'' wave is a commercial
Helium-Neon laser (Research Electro Optics model 30989, $\lambda=632.8$~nm).
The ``incoherent'' light source is a laser diode with a peak emission
wavelength at 650 nm, driven below the threshold current,  under which it
behaves as a light-emitting diode (LED).  In practice, there is no absolute
coherent or incoherent sources.  Rather, all light waves show a certain degree
of temporal coherence, given by the average correlation between the wave and a
delayed copy of itself. The parameter $\tau_{c}$ (correlation time) gives a
characteristic delay for which this correlation is maintained. Correspondingly,
the coherence length $\ell_{c}=c\tau_{c}$ is the distance the wave travels
during the time $\tau_{c}$. It is also worth noting that at the single photon level, this effect can be
related to the length or duration of the photon wavepacket.  The He-Ne laser
has a coherence length between 10 and 30~cm, when used in the operational regime. 
On the other hand, the spontaneous emission from the laser diode operated below threshold, due to its large bandwidth, has a reduced coherence length of less than 100 $\mu$m.

\begin{figure}
\centering \includegraphics[width=0.75\textwidth]{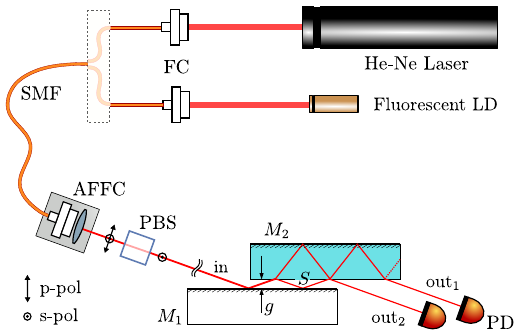}
 \caption{ Experimental setup. By alternatively coupling a single mode fiber (SMF) to the coherent or incoherent source, we select 
the corresponding probe beam. Light is later decoupled from the fiber 
using an adjustable focus fiber collimator (AFFC). The input polarization 
is selected with a polarizing beamsplitter cube (PBS). For this experiment 
we set the polarization state of the light perpendicular to the plane of 
incidence, using a polarizing beamsplitter cube (PBS) which reflects the 
orthogonal polarization away from the main beam. The two possible outputs 
are monitored with a pair of Si-biased photodetectors (PD); the intensity 
of the output, that propagates inside the glass is partially collected 
after an additional reflection in $M_2$, and a transmission through the 
interface $S$. 
The coherent light source is a He-Ne laser, while the incoherent light 
source is a laser diode (LD) pumped below the lasing threshold. Each beam 
is sent to a fixed-lens fiber coupler (FC). This allows us to select any 
of the light sources by plugging the fiber patchcord into the 
corresponding fiber coupler. 
The size of the gap $g$ controls the spatial overlap between the two beam 
paths.}
\label{fig:expset} 
\end{figure}

To ensure equal propagation conditions, each light source was sent to a multi-axis, high-resolution fiber
coupler (Thorlabs PAF-X-11-B) that allow light to be launched into
a Single Mode Fiber (SMF). Therefore, the input light beam for the
experiment was generated by coupling either the coherent or the incoherent
source to a 2 meter SMF patchcord and sending it to the experiment
using an adjustable focus fiber collimator (Thorlabs CFC-11X-B). The
experiment end of the fiber patchcord was left untouched throughout
the whole experiment, while the opposite end could be switched between
the two fiber couplers, to select either the coherent or the incoherent
light source. The purpose of the coupling and decoupling into a SMF
is twofold: to obtain good beam quality by spatial filtering with
the single transverse mode propagation of the fiber, and also to ensure
a similar alignment for both light sources (Figure \ref{fig:expset}).

Using the adjustable lens of the collimator, we loosely focus the
output beam down to a 400~$\mu$m waist at one meter from the lens,
where we placed the interference experiment. A polarizing beamsplitter
cube (Thorlabs PBS202) allows us to define a specific polarization for the input beam.

We use a 25~mm square, 6~mm thick N-BK7 borosilicate crown-glass
optical window as one of the propagating mediums ($n_{2}=1.515$).
Both surfaces are optically polished and one of them has an aluminum-coating
mirror deposited on it ($M_{2}$). The uncoated surface is the partially
reflecting interface $S$ between the BK7 glass and air. For the reflection
in air ($M_{2}$), we use the first surface of a circular, 25~mm
diameter, silver coated mirror. In this configuration, the optimum
incidence angle under ideal reflectivity conditions can be calculated
using the expression \eqref{Optimal}, which results in the value 
$\phi=78.57^{\circ}$, for $n_{1}=1$ and $n_{2}=1.5151$.

Both mirrors are mounted on precision angular mounts to ensure parallelism
between surfaces. Additionally, the silver mirror mount $M_{1}$
is placed on a manual, micrometer-resolution translation stage,
to allow for fine adjustment of the air gap $g$ between the optical
surfaces.  This controls the overlap between both beams. By laterally displacing the
two mirrors as depicted in figure \ref{fig:expset},
we gain access to the two possible outputs of the light,
after the interaction between the reflected and the transmitted beams.
The entire setup is mounted on a vibration-free optical 
table (STANDA Ltd.) to ensure that the interference pattern is not 
affected by mechanical vibrations or air currents.

\subsection{Preliminary characterization of surface reflectivities}

For a coherent wave, complete reversibility with a single interaction
can be expected, if the splitting ratio at the interface is close
to 50\% and there are similar losses on both paths. According to the
previous discussions, a reflectivity of $R_{S}=0.5$ can be obtained
at an interface between air ($n_{1}\approx$~1) and BK7 glass ($n_{2}=1.515$ for the range 
of wavelengths between 633 and 650 nm)
for an s-polarized field and an incidence angle as large as $78.6\degree$.
This is a fairly large incidence angle, and although
the transverse profile of the beam is small, it is difficult to avoid
partial blocking and distortions from the edges of the mirrors. Fortunately,
a mismatch between the reflectivities of mirrors $M_{1}$ and $M_{2}$
helps to avoid this problem. We measured the mirror reflectivites
for the actual experimental conditions: The silver mirror $M_{1}$
has a reflectivity of $R_{1}=0.91$ for an incidence angle $\theta_{i}\simeq75\degree$,
while the reflectivity of the aluminum mirror $M_2$, which is deposited at the opposite surface of the 
 BK7 slab $M_{2}$, was measured to be $R_{2}=0.61$ for an internal incidence angle of
$50\degree$. This is, to a good approximation, the complementary angle of the refracted ray inside the glass for
incidence angles between $74\degree$ and $78\degree$. The beam propagating inside the glass suffers an increased loss,
and therefore the incidence angle has to be adjusted in the experiment
to obtain a larger transmissivity, which compensates for this loss
imbalance, as discussed in Sec. \ref{realexp}.

\subsection{Balanced loss configuration}

\begin{figure}
\centering \includegraphics[width=0.66\textwidth]{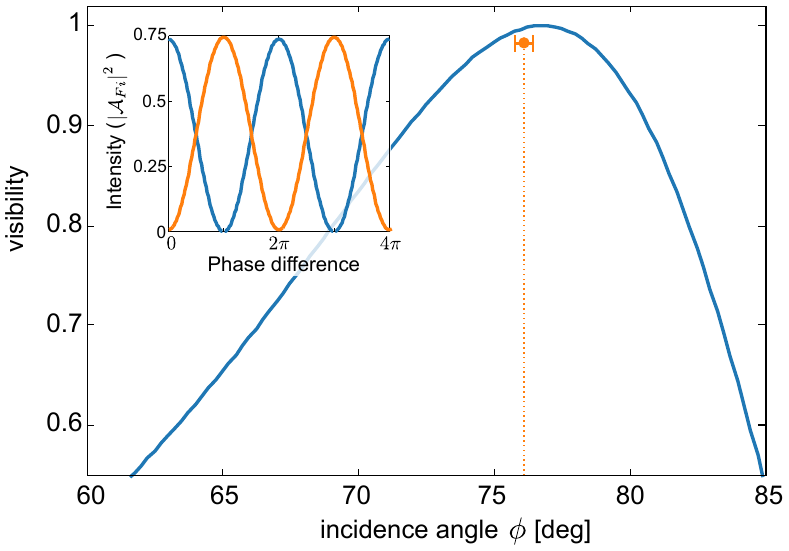}
\caption{Measured visibility as a function of incidence angle in
radians. The maximum is located at 1.34 radians, or $76.8\degree$. In
orange circular symbol is depicted the visibility value obtained experimentally.
It was measured for an incidence angle of $(76.1\pm0.6)\degree$.
The actual visibility is affected by several factors, other than the
balance of intensities, such as imperfections of the beam wavefront
and scattering at the surfaces. The
inset shows the calculated complementary modulation of the intensity at both outputs,
as a function of the phase difference between the two paths.
This calculation considers a perfectly coherent, unit intensity input beam. The maximum output intensity departs
from unity due to reflection losses.}
\label{fig:visibility} 
\end{figure}

Experimentally, we set the incidence angle at a value of $\theta_{i}=76.1\degree$
that maximizes the fringe visibility. Such angle leads to a ratio
between transmissivity and reflectivity given by the Fresnel coefficients
of $T_{S}/R_{S}=1.31$ and a refraction angle $\theta_{r}=39.8\degree$.
Substituting the measured reflectivities $R_{1}$ and $R_{2}$ in to
the expression \eqref{Antiphase2}, we find that the theoretical ideal incidence 
angle would be $76.8\degree$. Remarkably, this is very close to the experimental 
value of $76.1\degree$ (See Fig. \ref{fig:visibility}).

For such configuration, an air gap of a nominal thickness of 1.24~mm
equalizes the distance traveled by both beams in the direction parallel
to the interface. In this situation, depending on the coherence of
the source, total recombination of the beams into a single beam may
be obtained at the output. We point out that this must be
understood as a signature of reversibility. Furthermore, the light
can be arranged to exit the experimental
setup as a single beam through either of the outputs. Depending
on the phase difference between the two paths, the coherent combination
of the beams will lead to constructive interference alternating between 
 each of the  outputs.  This is shown in the inset of figure \ref{fig:visibility}.

\section{Results \label{sec:Results}}

After a careful adjustment of the separation $g$ (see Fig. \ref{fig:expset})  and a
careful mirror alignment, the coherent source gives rise to a complementary single-fringe interference
pattern,can be observed in the two outputs. We use two large-area biased Si photodiodes
(Thorlabs DET-36A) to register the light intensity at both outputs.
The output that propagates in air is fully collected by the detector,
while the one that propagates inside the glass is attenuated due to
an internal reflection in $M_{2}$, and a partial transmission from glass to air at the interface.
This reduces the intensity by an overall factor
of $R_{2}T_{S}=0.346$. The relative responsivity of the 
two photodiodes was measured by directly comparing their outputs under 
identical illumination conditions, using an attenuated 633 nm laser beam 
with a spot size of 1~mm (1/$e$ diameter), measured at the position of the 
sensors. The relative difference in responsivity was found to be below~1.5~\%.

\begin{figure}[H]
\centering \includegraphics[width=0.75\textwidth]{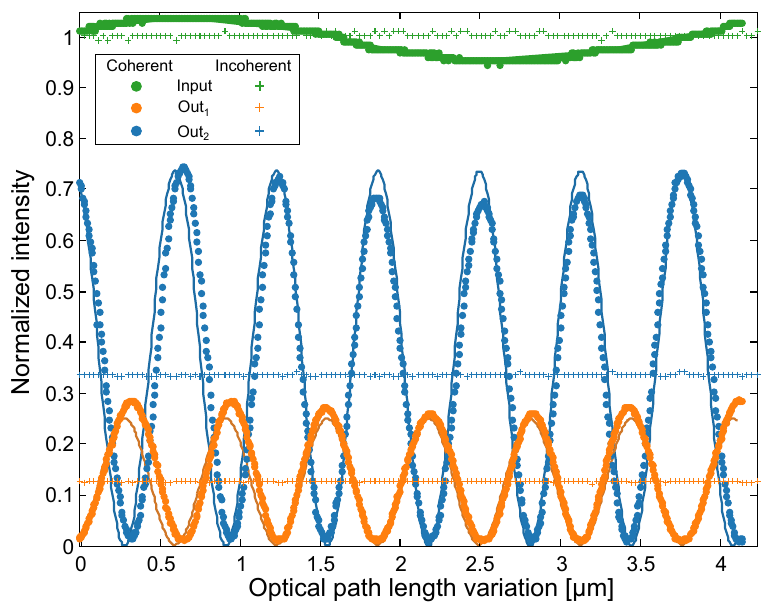} \caption{Measured intensities at the input and at the two complementary outputs
for both, coherent and non-coherent light sources, as a function of the variation of the optical path length.
The modulation was generated by changing the air gap width between
the two mirrors. The coherent source shows almost ideal time reversibility
when losses at the mirrors are taken into account: all the light left
in the system is combined into a
single beam (full circles). In contrast, when the incoherent source is used no
modulation is present (plus symbols). This shows that the recombination of the two non-coherent
beams always gives one result, which is half the maximum possible intensity for the coherent
beams. In thin continuous
lines are the theoretical predictions are shown. These were obtained by
 using the experimentally determined incidence angles
and mirrors
reflectivities
measured experimentally.
    }
\label{fig:results} 
\end{figure}

Figure \ref{fig:results} shows the normalized intensities obtained experimentally, for the
input and output beams 1 and 2, as a function of the optical path
length variation (OPD), obtained by changing the size of the air gap
$g$ between the two mirrors. When the coherent source is used, the
relative phase between the two beams produces a complete recombination
of the light in a single beam, which exits the experiment through
either of the outputs.  The coherent light signals are shown with full circles in Fig. \ref{fig:results}.
For OPD differences that are integers of the light wavelength ($N\lambda$,
$N\in\mathbf{N}$), all the available light exits through the air
side of the interface (Out$_{2}$ in Fig. \ref{fig:expset}), while for path differences that are half integers
of the wavelength $(N+1/2)\lambda$, the light is almost completely
recombined on a single beam at the glass side (Out$_{1}$). The fringe visibility,
calculated using Eq. \eqref{Visibility} is above 97.5\% for both
outputs.

Furthermore, taking into account the additional reflections and refractions
suffered by both outputs, the ideal \emph{mean} throughput at each
output can be calculated as: 
\begin{equation}
\begin{aligned} \langle \textrm{Out}_{1} \rangle  & =R_{1}\left(R_{S}T_{S}R_{1}+R_{S}T_{S}R_{2}\right)R_{2}T_{S}=0.118,\\
  \langle \textrm{Out}_{2} \rangle  & =R_{1}\left({R_{S}}^{2}R_{1}+{T_{S}}^{2}R_{2}\right)=0.334.
\end{aligned}
\label{eq:throughput}
\end{equation}
These values, averaged over the phase differences, exhibit an excellent agreement with the normalized intensities obtained using the incoherent source, and they represent the mid-point of the intensity excursions obtained with the coherent light source. That aligns perfectly with our derivation of the formulas for non-coherent light by averaging over the phase differences between paths \eqref{Intencity}.

When the experiment is carried out using the incoherent
light source, no intensity modulation is observed, and there is an even distribution
of light at the outputs, regardless of the phase condition given by
the variation of the OPD. 

For comparison, the theoretical predictions for the intensities at
both outputs are also shown in full lines, presenting a very good agreeement with the experimental data.
These are shown with plus symbols in Fig. \ref{fig:results}.

Figure \ref{fig:results} also shows a slowly varying intensity modulation 
in the coherent source (He-Ne laser), observed both at the input and at the output signals. 
This effect is caused by polarization modulation due to thermal drift in the laser cavity 
\cite{niebauer1988frequency}. A single period of this modulation can be observed as we
scan the air gap size.

\section{Discussion  \label{sec:Discussion}}

\subsection{Various cases of non-coherence}

It is important to mention that, although this experiment has been carried
out with intense light sources, similar results can be obtained at
the single photon level. The temporal coherence, namely the length
of the photon wavepacket, scales as the inverse of its spectral dispersion. The reversibility of the splitting at the interface within the first
interaction is related to the interference at the second interaction:
the fringe visibility will survive at full strength (reversible process)
if the optical path difference is much smaller than the coherence
length of the photon \cite{loudon2000quantum}. If the path difference
exceeds the duration of the photon wavepacket, both contributions
of the photon at the transmitted and reflected superposition state
will arrive separately at the interface and the interference vanishes.

In this single-photon picture the matrices $\mathcal{T},\,\mathcal{U}$
have a different interpretation. The elements of $\mathcal{T}$ may be interpreted
as the probabilities of photon transmitting through the interface
and the elements of $\mathcal{U}$ as probability amplitudes of transmission
through the interface. The difference between the coherent and incoherent
regimes is understood in this case, as the regime where the transition of a photon is described
by probability amplitudes with specific phase values and the regime
where only the absolute value of the probability amplitude matters.

Another possible setup to perform the experiment, like the one that we present here, is an experiment with two different light sources.  However, it is well-known that two independent  light-sources cannot exhibit stable interference. In a different set-up, the interference of two lasers was studied many times starting from Ref.~\cite{magyar1963interference}. For a more modern experiment see Ref.~\cite{kawalec2021observation}. While interference patterns form at any given moment, over a time interval comparable to the decoherence time, the interference pattern drifts: constructive interference may turn into destructive interference and vice versa. Therefore, when calculating the mean intensity over a time interval much greater than the coherence time, it should be averaged over phases.

We want to add that it is possible to obtain a stable interference of two separate but synchronized lasers \cite{salomon1988laser, zhang2024monolithic}. However, when synchronized, they are not two independent light sources.

 A mathematical explanation of the impossibility of forming a stable interference of independent light sources may be the following. If we substitute $|A|^2 = |A_1 + A_2|^2$ formula \eqref{fullint} and expand the bracets, the term that correspond to the interference of the two sources takes the form
\begin{align}
\text{Int} =\frac{1}{\Delta\phi} \int \limits_0^{\Delta\phi} d\phi A_1^*(\phi )  A_2(\phi ) + C.C.
\label{SumInt}
\end{align}
The integral expression defines a scalar product in an infinite-dimensional Hilbert space. In $n$-dimensional space, the expected value of the scalar product of two random unit vectors is proportional to $n^{-1/2}$ \cite{vershynin2018high}. So as $n \to \infty$, it tends to zero. This means that if we pick two random sources, their interference defined by formula \eqref{SumInt} is zero.

The single-source experiment that we have conducted is particularly interesting because it allows for direct experimental measurements in both coherent and non-coherent regimes, and the transition between these regimes can be described theoretically. However, we suppose that a two-source non-coherence is the direct analog of the non-coherence transformations of particles incident at the interface during the heat transfer.

Now that we have described the experiment and its possible modifications, let us reflect on how different the non-coherence concept that we use is from the one used in decoherence theory and quantum thermodynamics. In both theories, the decoherence is the result of the interaction with the environment. In our experiment, the coherence disappears without any interaction with the medium: there is no photon scattering, collisions or any other interaction, due to the nature of the optical materials used and the photon cross section. Non-coherence is a result of the difference in the traveled distances between the two parts of the beam. Nothing affects the beam while it is traveling through nonabsorbing media. If we let each of the separated parts of the light travel an arbitrary long but equal optical distance, before putting them back together, they will remain mutually coherent.  

Moreover, in the conducted experiment, there is indeed a process that might be considered as analogous to interaction with an environment. It is the interaction between the light and the mirrors. As it was pointed out, a considerable portion of light was absorbed (or transmitted) rather tha reflected by the mirrors. However, this process does only affects the intensity of the field (i.e. the mean photon number of each beam) and eventuallly the global phase of the beam, but not the phase of each photon. Therefore, no loss of coherence is produced at the reflections. The incidence angle of the light, which controls the portion of reflected and transmitted light via the Fresnel equations, was fine tuned to compensate for differences in the transmission of each beam, before the second interaction at the interface in order to combine two fields of approximately equal amplitudes.

 The experiment demonstrated the reversibility of the process of transmission/reflection of coherent light beams at the interface between the air and the glass and the non-reversibility of that same process for non-coherent light beams. In both cases, the interaction with mirrors was equally strong, which means the same portion of light was absorbed. That interaction did not affect the result of the experiment even slightly; coherent light beams remained coherent and, under appropriate phase conditions, they were successfully merged back into a single beam.

\subsection{Thermo-optical analogy}

We interpret our optical experiment as an analogous
to the process of heat transport through an interface. Such analogy
allows for studying and understanding the emergence of irreversibility
in statistical-mechanics systems, in the powerful and clean framework
of optical physics. If we examine different types of quantum particles
such as phonons, electrons, and photons, transitioning through the
interface, the equations that connect amplitudes of incident waves
and departing waves take, universally, the form \cite{meilakhs2024transmission}
\begin{align}
\begin{pmatrix}A_{D1}\\
A_{D2}
\end{pmatrix}=\mathcal{U}\begin{pmatrix}A_{I1}\\
A_{I2}
\end{pmatrix},\label{Universal}
\end{align}
where $\mathcal{U}$ is a unitary matrix. The expressions for the
elements of $\mathcal{U}$ are different for the different types of
particles and the different models, but the unitarity is universal.

Starting from this equation, we derive another expression, that
relates the distribution functions of particles departing from the
interface with the distribution functions of particles incident on
the interface 
\begin{align}
\begin{pmatrix}f_{D1}\\
f_{D2}
\end{pmatrix}=\mathcal{T}\begin{pmatrix}f_{I1}\\
f_{I2}
\end{pmatrix}.\label{Matching}
\end{align}
Distribution functions $f$-s are defined as the mean number of particles
in the given state, $\mathcal{T}$ is a matrix in which the elements
are probabilities of transmission or reflection of particles at the
interface. The elements of $\mathcal{T}$ are calculated as squares
of absolute values of the elements of $\mathcal{U}$ from equation
\eqref{Universal}. We call this equation a matching equation for
the distribution functions.

To calculate the Kapitza resistance, we accompany the matching equations with
thermalization conditions on each interface side. These conditions are needed
to connect the distribution functions of particles with the temperatures at
both sides of the interface. These differ by the value of the Kapitza
temperature jump. In the setting of heat transport, the distribution functions
of particles are not given directly, but they should be expressed as functions of
temperatures. In an optical setting, however, we do not have physically
meaningful analogous conditions. Instead, we point out that the intensity of
light is the mean number of photons multiplied by the flux associated with each
photon, $I= f c \hbar \omega $, which  means that the distribution of particles can be 
measured directly. Therefore, we only replicate the matching equations in the
optical experiment.

We obtain the matching equations by finding an expression for the
matrix $\mathcal{T}$. To this end, we need to find an explicit form
for matrix $\mathcal{U}$. The natural way to find $\mathcal{U}$
is by solving the following problem: let the particle be incident
on the interface from one side with probability one, and find the
probability amplitudes of its transmission and reflection. The solution
for the case of incidence from a side $1$, for example, gives
the values for the first column of $\mathcal{U}$, while the solution
for the case of incidence from side
2, gives the values of the second column (see Fig. \ref{fig:genscheme}). This physical
situation can be also stated in a different way. Let us assume that
some particle is departing from the interface on the given side with
certainty, and find the probability amplitudes of its incidence from
each side. By solving this problem for both sides of departure, we
get the equation 
\begin{align}
\begin{pmatrix}A_{I1}\\
A_{I2}
\end{pmatrix}=\mathcal{U}^{\dag}\begin{pmatrix}A_{D1}\\
A_{D2}
\end{pmatrix},\label{Inverse}
\end{align}
which is mathematically equivalent to the equation \eqref{Universal}.
Indeed, it
can be obtained from it, by multiplying both sides of Eq. \eqref{Universal}
by $\mathcal{U}^{\dag}$.

Intuitively, the equation \eqref{Inverse} is a time-reversed version
of equation \eqref{Universal}, since the latter describes the splitting
of waves at the interface, and the former the merging of two waves
at the interface. Since wave equations are time-symmetric, both are
equivalent. If we apply to Eq. \eqref{Inverse} the same procedure
that we
used for \eqref{Universal}, we obtain the following matching equation
for the distribution functions 
\begin{align}
\begin{pmatrix}f_{I1}\\
f_{I2}
\end{pmatrix}=\mathcal{T}\begin{pmatrix}f_{D1}\\
f_{D2}
\end{pmatrix}.\label{ReMatching}
\end{align}
By solving these equations with the same thermalization conditions,
we find that for the same value of temperature difference between
sides, we get the opposite values of heat flux. This means that heat
flows from the colder body to the hotter body. We indeed made the
time reverse.

However, equations \eqref{Matching} are not equivalent to \eqref{ReMatching},
since $\mathcal{T}\neq\mathcal{T}^{-1}$. We started with two equivalent
equations,namely Eqs. \ref{Universal} and \ref{Inverse}, applied
the procedure to them, and obtained two non-equivalent equations,
i.e. Eqs. \ref{Matching} and \ref{ReMatching}. Thus we have lost
reversibility during the derivation of the matching equation.

In Ref. \cite{meilakhs2024transmission} was conjectured that the
juncture in the derivation where the time symmetry is disrupted, is
the assumption of the absence of phase correlation between particles
that are incident on the interface from opposite sides. That allows
for averaging over amplitude phases which, in turn, allows for the
description of processes in terms of probabilities and their bistochastic
transformations \eqref{Bistoch} instead of amplitudes and their unitary
transformations \eqref{Unitary}. As we have pointed out before, bistochastic
matrices (ingnoring exceptional cases) do not have inverses, which
means that they represent irreversible transformations.

The presented experimental results entirely confirm this suggestion.
In the coherent regime the phase information is present, complete,
and can be measured exactly, and we have found perfect reversibility.
In constrast, for the non-coherent regime all the information about
phases is lost and, indeed, we found that this state is irreversible.
This is proved because the transformations of intensities at the interface
are successfully described by bistochastic matrices \eqref{Bistoch},
obtained from unitary matrices by squaring absolute values of their
elements. Since intensities $I$ and the mean number of particles
$f$ are connected by formula $I=fc\hbar\omega$, they transform identically.
 This is exactly what we observe
by comparing \eqref{Non-coherent1} with \eqref{Matching} and bearing
in mind the former is a two-step transformation.

For this experiment, we have specifically employed light sources with
close wavelengths and intensities (see Section III), and the only
feature that accounts for the qualitative difference in the results
is the coherence length of the sources.

The main difference between the experimental setting studied in this
work and the case of heat transport is that the former allows to explore
both, the coherent and the incoherent regimes. In a heat transport
setup, the indistinguishability condition that eventually leads to
interference (i.e. coherence) and reversibility is very hard to achieve.
This condition involves the interaction of particles that are incident
on the interface from opposite sides and hence cannot have phase correlation. See the discussion of two-source interference in the previous sub-section.  As a consequence, heat transport occurs
permanently in a non-coherent regime and it is always irreversible.

\section{Summary}

In this work we have carried out an experiment
that shows a strong connection between irreversibility and coherence
and provide a bridge between the frameworks statistical mechanics
and optical physics. In the experiment, two parts of a light beam, initially split upon
incidence at the interface, are redirected to the interface and meet at a common spot. In
the condition such that the difference between the lengths of their optical paths is much
greater than the coherence length of light (non-coherent regime),
the intensities measured on the outputs agree with those calculated
by the formula \eqref{Non-coherent2}. In the opposite case, when
the difference between the lengths of optical paths was much shorter
than the coherence length of light (coherent regime), the measured
intensities that agree with those calculated by the formula \eqref{Coherent2}.
In the latter case, by varying the difference between the lengths
of optical paths, we were able to combine the split parts of the light
back into a single beam, which confirms the reversibility of the process
in the coherent regime. The non-coherent regime, on the contrary,
does not allow for such a recombination, thus producing an irreversible
process.

We hope that this powerful analogy between coherence and reversibility,
and our successful mapping from the physics of heat transfer to that
of optics of beams incident in an interface, can provide a bridge
between these two research areas and motivate further experimental
and theoretical work. 

 While we have established with sufficient confidence that Kapitza resistance arises due to the non-coherence of incident waves, the next goal is to explore situations in which semi-coherent phenomena may occur. We suggest that transport in nanostructures offers a promising opportunity, since the characteristic scale of these systems may be smaller than or comparable to the coherence length. The essential task would be to vary system parameters in order to observe the transition from coherent to non-coherent transport, under conditions where the particle mean free path remains much larger than the structural scale. Demonstrating such a transition would establish that the mean free path and the coherence length are distinct physical quantities, thereby providing unambiguous evidence that the loss of coherence cannot be attributed to interactions with the environment.

Transport in nanostructures is a rapidly developing field \cite{badawy2024electronic, shiri2024low, mangnus2022charge}, and the distinction between coherent and non-coherent transmission has already been discussed in relation to interfaces \cite{chen2011phonon, chen2024thermal}. It is therefore quite possible that some effects consistent with the present viewpoint have already been reported in the literature.

Summing up, we suggest that an important direction of investigation, which develops the proposed analogy between optics and heat transfer, is the study of systems that are neither fully coherent nor fully non-coherent -- systems in which the coherence length must explicitly enter the results of calculations. This is equally relevant in the context of optics and in the context of transport theories at the nanoscale.

Other important route to explore is extending the scope of this analogy from interfaces to bulk materials, which can ultimately shed light on the general arrow of time problem.

\bibliographystyle{ieeetr}
\bibliography{thebilio,timerev,Kapitza}

\end{document}